# Research of migration behavior of space charge packet in polyethylene by electron beam irradiation method under the applied electric field

**Hui Zhao[2], Yewen Zhang[1, 2]*, Jia Meng[2], Feihu Zheng[1] and Zhenlian An[1]**

[1] Department of Electrical Engineering, Tongji University, Shanghai 201804, China

[2] Shanghai Key Laboratory of Special Artificial Microstructure Materials and Technology, Department of Physics, Tongji University, Shanghai 200092, China

(Email: yewen.zhang@tongji.edu.cn)

**Abstract:**

For accurately obtaining the relationship between the carrier mobility and the applied electric field, a new multi-layer sample has been designed. Polyvinyl fluoride (PVF) films were hot pressed on both sides of linear low density polyethylene( LLDPE) to block the charge injection from the electrode, so as to better observe the migration of irradiated electrons. The new multi-layer sample was firstly charged to form a charge packet in the electron beam (e-beam) irradiation setup. And then it was transferred to the Laser Induced Pressure Propagation (LIPP) setup to have the space charge evolution monitored under DC voltages on the order 10-70 kV/mm. The migration of the charge packet has been successfully obtained in this new multi-layer sample. By using the packet front as the reference point, the range of the average mobility of packets at a range from $0.06\times10^{-14}$ to $1.02\times10^{-14}$ $m^2/(V\cdot s)$ under calibration local field. The experimental results coincide well with the curve relating charge mobility and the electric field predicted from the Gunn effect-like model.

## 1. INTRODUCTION

Charge carrier mobility is a very important parameter for dielectric materials[1, 2] especially for the widely adopted polymeric insulations[3]. Although many researchers presented various methods to measuring carrier mobility in dielectric matetials, the currently accepted feasible measurements are as follows: time-of-flight method[4-6], transient current method[7, 8], surface potential decay method[9, 10] and irridiation eletron method[11, 12].

Recently, various models have been suggested to give a reasonable explanation of the packet-like space charge behavior which has been found in polyethylene material. A Gunn effect-like model, being proposed by many researchers[13-15], postulates that there exists a negative differential relationship for the charge velocity against the electric field when the electric field exceeds a threshold value. Based on this model, the simulation was demonstrated to be able to fit the experimental results very well for different cases[16]. However, it is impossible to verify the existence of the Gunn effect directly from the band structure point of view because of the complicated macrostructure of polymeric materials (e.g. crystalline regions, amorphous regions and the interfaces among them).

In this paper, an electron beam irradiation system was constructed for injecting electrons into the multi-layered structure of polyethylene samples. By applying an accelerating voltage on the upper electrode of the electron beam irradiation instrument, amounts of electrons are initially injected into the sample at a certain depth from the surface to form a charge carrier packet. It has to be pointed out that the obtained carrier mobility is valid only when the external electric field is much larger than the electric field resulting from the charge packet itself. Then an external electric field is applied on the sample and the charge carrier packet starts to move. The apparent charge carrier mobility can thus be obtained from the carrier packet velocity under the electric field.The relationship between the carrier packet velocity and the applied electric field was obtained by LIPP method to test the validity of the Gunn effect-like model.

## 2. SAMPLES PREPARATION AND EXPERIMENT

The raw material of polymer samples used in this study is the granular product LL1004 from Exxon Mobile Chemical C. Its characteristics are as follows: density of 0.918 g/cm$^3$ and melt index of 2.8g/10min. The neat LLDPE particles were pressed by flat vulcanizing machine into the sheet samples at 130°C under the pressure of 15MPa for 12 min with a size of 0.5 mm in thickness and of 170 mm in diameter. PVF film with a size of 25μm in thickness and of 60mm in diameter was hot-pressed onto both sides of the sheet samples as blocking layers, and then both sides of the samples were evaporated by aluminum (Al) electrodes of 50 mm in diameter which can guarantee the reliable grounding when the samples are irradiated by the e-beam. The average thickness of Al electrodes is approximate 20nm. The structure of sample with blocking layers and Al electrode is shown in Fig. 1.

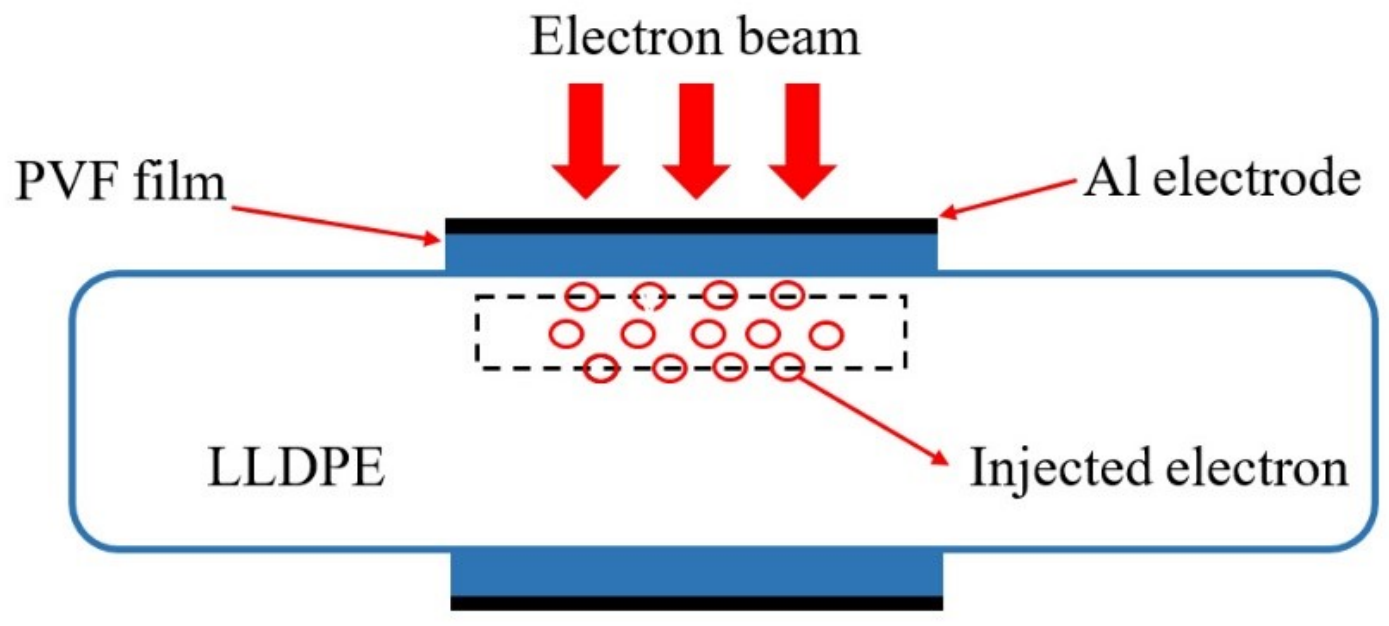

Fig. 1 Irradiated LLDPE sample hot-pressed with PVF films

The schematic setup for electron beam irradiation is as shown in Fig. 2. A filament used for the scanning electron microscope(SEM) acts as the electron beam source. The emitted electrons are directly accelerated toward the sample under a certain dc high voltage which can be adjusted between 0 kV and 80 kV. Without adjustment of e-beam, the quantity of injected charges in sample with higher accelerated energy is relatively larger than that by lower accelerated energy for the same irradiation duration. The sample with a multi-layered structure was placed at a grounded Al substrate with a diameter of 200 mm to ensure the bottom electrode reliable grounding. For the same purpose, a grounded Al plate with a concentric hole 40mm in diameter to allow the samples exposed to the electron beam. In this study, the accelerated voltage is selected for 70kV, the irradiation lasted for 180 s with current density 1.8μA/cm$^2$ in a vacuum of 7.5×10$^{-4}$ Pa. After the irradiation, the Al electrodes on both sides of the sample are covered with the poly(ethylene-co-vinyl acetate)(EVA) electrodes filled with carbon black particles . The semiconducting electrode was prepared by hot pressing the EVA granules at 90°C under the pressure of 15MPa for 6min with a thickness of about 0.4 mm and of 50 mm in diameter. In order to help propagation of pressure waves signal at interfaces between electrodes and the sample, A semi-conductive layer (silicone oil) should be placed between the sample and the Al electrode to improve the acoustic impedance matching.

The irradiated multi-layered sample is transferred into the LIPP system for space charge measurement. This system is composed of the laser (Continuum SureliteII-10, wavelength of 1064 nm, laser pulse width of 7ns, and power of 500mJ), digital oscilloscope (Tektronix TSD5052), and a wide band amplifier (Miteq, 37dB). The principle and details of the LIPP system can be found in previous numerous publications[17, 18].

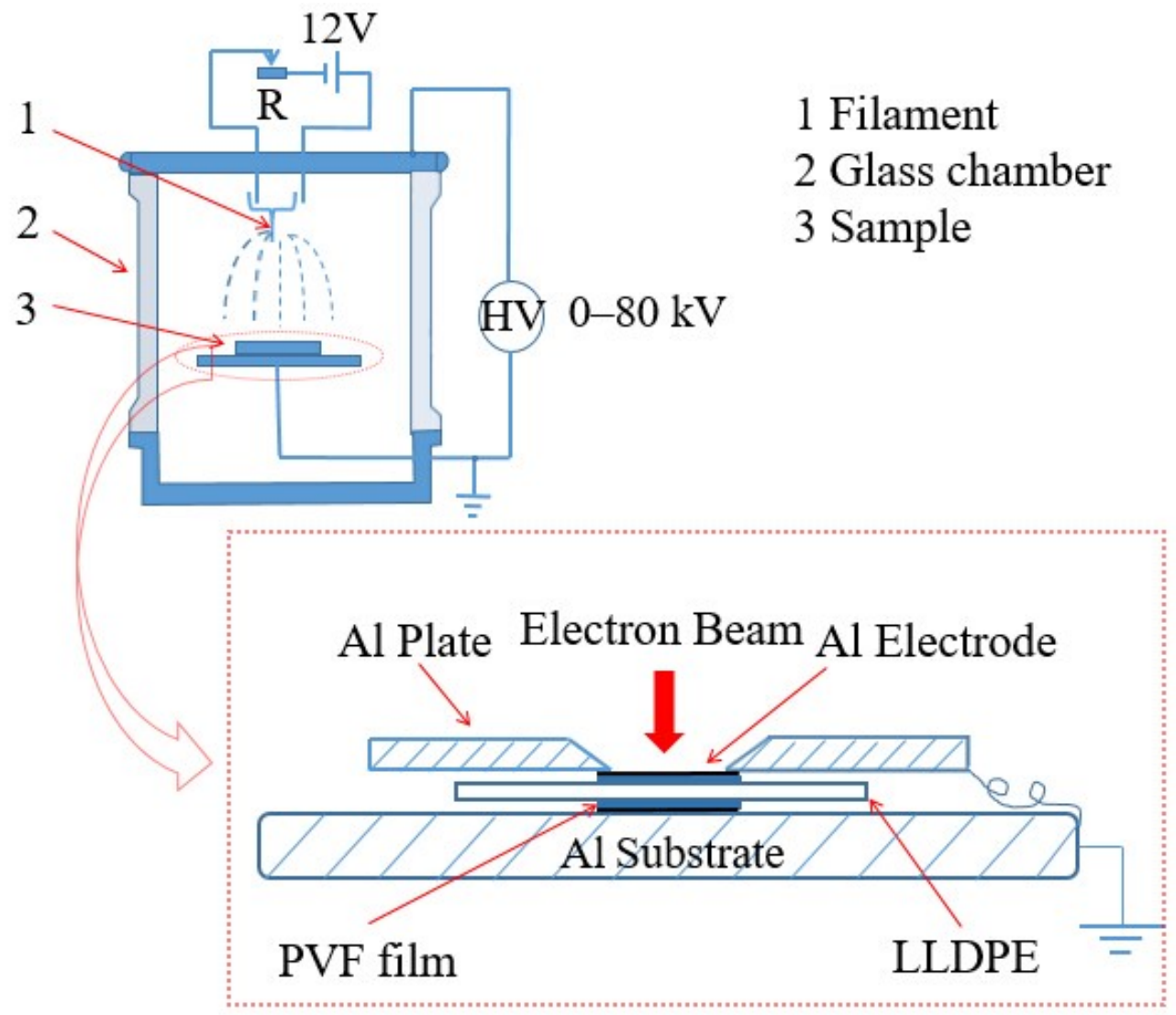

Fig. 2. Schematic setup for electron beam irradiation.

## 3. Experimental Results and discussion

Figure 3 is a typical experimental result of the multi-layered sample polarized at 30 kV/mm DC voltage at 25°C. As is shown in the figure 3(a), a broad negative charge packet with a peak was detected in the bulk when measurement started, it's noticeable that the injected electron exists in the whole irradiated region. The maximum range of the injected electron located near the irradiated electrode with the distance of 80μm.

Concerning about the penetration depth of the injected electron, the electron penetration depth can be calculated by an empirical formula suggested by Weber [19]:

$$R = \frac{\alpha E_0}{\rho}\left(1 - \frac{\beta}{1+\gamma E_0}\right)\ \text{cm} \qquad (1)$$

Where, $\alpha$=0.55 g/cm$^2$/MeV, $E_0$ is the electron energy in MeV, $\rho$ is the dielectric density, $\beta$=0.9841, $\gamma$=3 MeV$^{-1}$. According to Eq. (1), the electron beam energy of 70keV correspond to the electron penetration depth are about 80μm, which is in good accordance with the experimental results.

These injected electrons formed a layer of negative charges that can migrate to the opposite electrode under a certain driving voltage. Corresponding to the experimental results, the space charge packet can be observed by the LIPP method. A typical space charge packet was obtained. After the measurement commenced, the negative charge packet moved from the negative electrode side to the positive one. As it moved on, the height of the peak of the packet gradually decreased. At the meantime, both positive and negative homo charge injections were invisible, denoting the well charge blocking effect of the PVF films[20]. According to the Gunn effect-like model, the velocity of charge packet will get slow during the transportation. This phenomenon is also verified by the experimental results as shown in the figure 3(a).

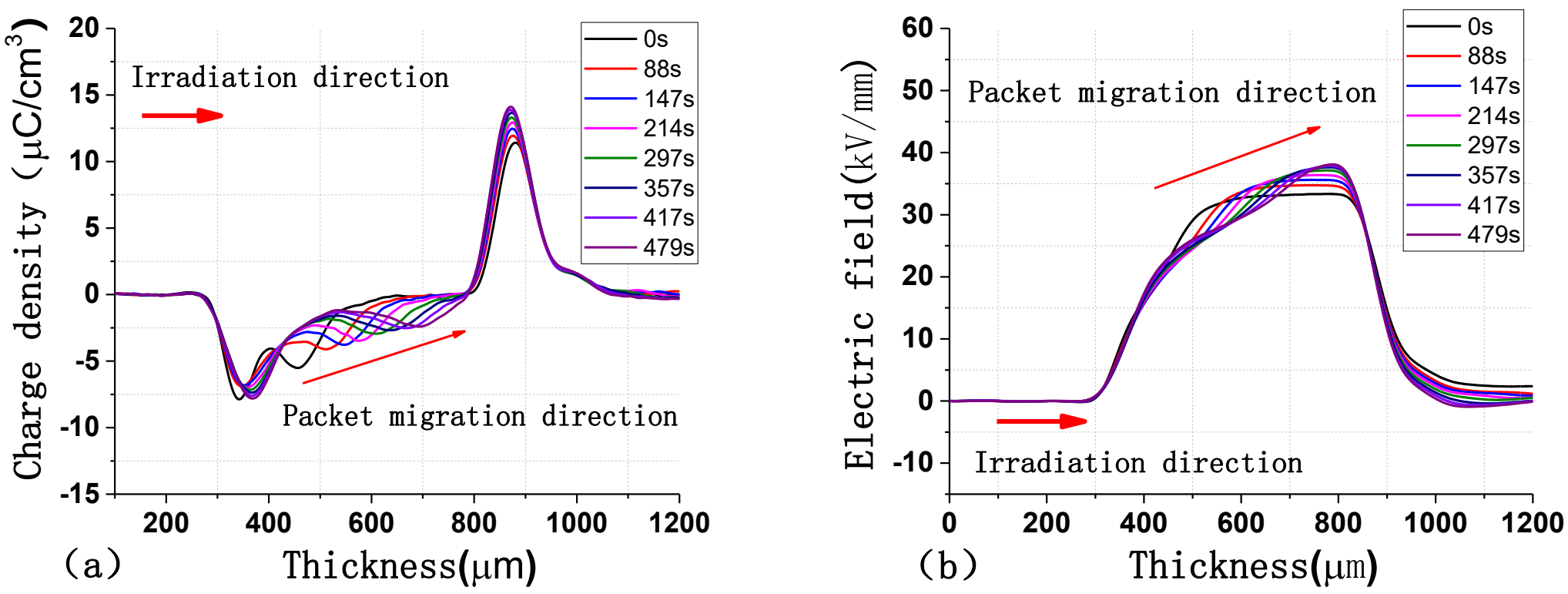

Fig. 3. Typical results of charge distribution and calculated electric field profiles in a multi-layered sample under 30 kV/mm

Figure 3(b) show the electric field distribution calculated from the charge profiles shown in figure 3(a). Actually, it must be stressed that the field distribution is no longer exactly the capacitive field distribution as the existence of space charge in the process of charge packet migration. The calculated results indicate that the captured charge, together with the charge packet, induce a significant distortion of local field. The field near the cathode keeps the same value no matter where the packet charge locates in, which means the PVF films own the excellent blocking effect as above mentioned. The local field gradually increases up to the rear edge of the packet and is sharply enhanced in the position of charge packet observed. The field at the front of the packet increases during the process of charge packet migration.

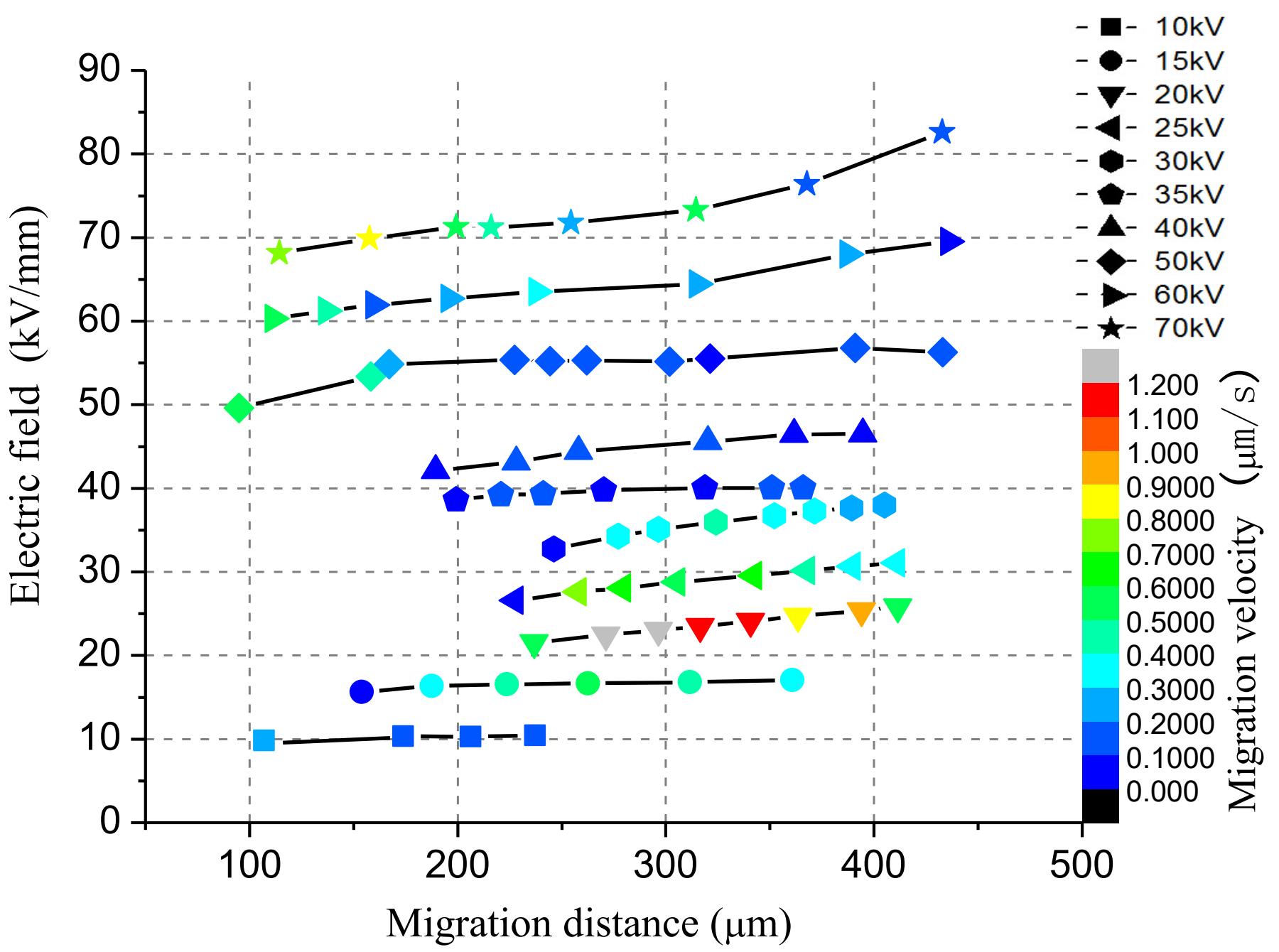

Fig. 4. Dependence of calibration local field on migration distance

Figure 4 shows the dependence of calibration local field on migration distance under each applied DC field from 10kV/mm to 70kV/mm. The colour scale and numbers to the right represent the migration velocity of charge packets in μm/s (same for all plots). Two phenomena are worth mentioning here. On the one hand, for the X axis, all the migration velocity of packet possesses the gradually decreasing trend for each sample under different applied field, which has confirmed by the above experimental results. It's also consistent with the inference of the Gunn effect-like model. On the other hand, for the Y axis, the maximum value of the average migration velocity approximately appeared at the 25 kV/mm, it's about $1.02\times10^{-14}m^2/(V\cdot s)$. Correspondingly, the minimum value existed little beyond the 40 kV/mm, it's about $0.06\times10^{-14}m^2/(V\cdot s)$. This phenomenon is also verified by the following analytical results as shown in the figure 5.

As shown in the Figure 5, the red curve represents the average velocity of charge packets, this figure also shows the dependence of packets velocity on calibration local field for each stressing sample from 10kV/mm to 70kV/mm. The colour scale and numbers to the right represent the migration distance of charge packets in μm (same for all plots). According to this figure, we analysed from two aspects. Firstly, for each applied field, the migration velocity of charge packet decreases apparently along its propagation through the insulation which is consistent with the earlier analytical result. Secondly, according to the red curve, negative differential resistance can be observed and the onset of the negative differential resistance was found to be approximately as low as 25kV/mm and the second inflection point appeared little beyond 40kV/mm. The shape of this curve is strikingly similar to the Gunn effect occurring in semiconductors. Based on the analyses above, the experimental results coincide well with the curve relating charge mobility and the electric field predicted from the Gunn effect-like model.

In this experiment, a range of electric fields from 10 kV/mm to 70 kV/mm were applied to irradiated samples and the average velocity of charge packets were obtained by the LIPP method. When the applied field reaches to 70 kV/mm, the curve of the average velocity was able to fully demonstrate the whole extent of the problem. Therefore, the present results are limited to 70 kV/mm. According to the propagation process, repetitive charge packets were not observed, this is because the source of charge packets is the irradiated electron, and the existence of the PVF film effectively blocked the injection of space charge from the sample electrode, but charge fronts propagation as relatively wave-packet could be identified obviously.

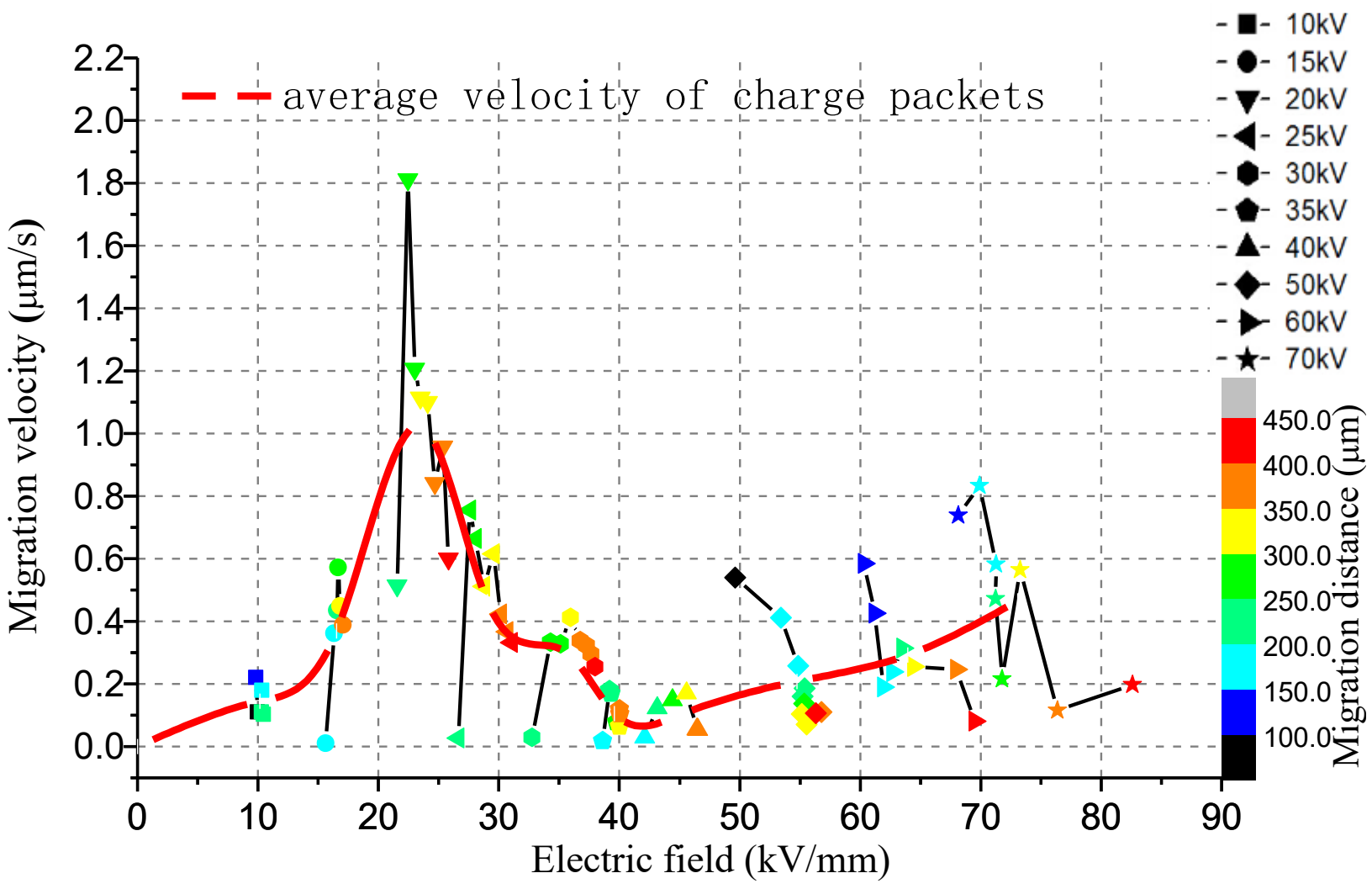

Fig. 5. Dependence of packets velocity on calibration local field

## 4. Conclusion

The dynamics of charge packets in the new multi-layer polyethylene samples have been investigated through LIPP method under various DC electric fields. In this paper, the source of charge packets is the irradiated electron which is ejected by the electron beam irradiation setup. According to the calculation results, the range of the average mobility of packets at a range from $0.06\times10^{-14}$ to $1.02\times10^{-14}$ $m^2/(V\cdot s)$ under calibration local field and the results coincide well with the curve relating charge mobility and the electric field predicted from the Gunn effect-like model.

## 5. Acknowledgment

Financial support from the National Natural Science Foundation of China (NSFC Nos. 51277133, 51477118) and Ph.D. Programs Foundation of Ministry of Education of China (No. 20130072110046) is gratefully acknowledged.